\documentclass[%
reprint,
amsmath,amssymb,
aps,
]{achemso}

\usepackage{caption}
  
\usepackage{subcaption}
\usepackage{graphicx}
\usepackage[]{babel}
\usepackage{braket}
\usepackage{tikz}
\usetikzlibrary{decorations.pathmorphing}
\usetikzlibrary{shadows}
\usepackage[utf8]{inputenc}
\usepackage{hyperref}
\usepackage{xcolor}

\title{Theory of hot-carrier generation in bimetallic plasmonic catalysts}

\author{Hanwen Jin}
\affiliation{Department of Materials, Imperial College London, South Kensington Campus, London SW7 2AZ, UK}

\author{Matias Herran}
\affiliation{Nanoinstitute Munich, Faculty of Physics, Ludwigs-Maximilians-Universit\"at M\"unchen, 80539 Munich, Germany}

\author{Emiliano Cort\'es}
\affiliation{Nanoinstitute Munich, Faculty of Physics, Ludwigs-Maximilians-Universit\"at M\"unchen, 80539 Munich, Germany}

\author{Johannes Lischner}
\affiliation{Department of Materials and the Thomas Young Centre for Theory and Simulation of Materials, Imperial College London, South Kensington Campus, London SW7 2AZ, UK}
\email{j.lischner@imperial.ac.uk}

\begin{document}

\begin{abstract}
Bimetallic nanoreactors in which a plasmonic metal is used to funnel solar energy towards a catalytic metal have recently been studied experimentally, but a detailed theoretical understanding of these systems is lacking. Here, we present theoretical results of hot-carrier 
generation rates of different Au-Pd nanoarchitectures. In particular, we study spherical core-shell nanoparticles with a Au core and a Pd shell as well as antenna-reactor systems consisting of a large Au nanoparticle with acts as antenna and a smaller Pd satellite nanoparticle separated by a gap. In addition, we investigate an antenna-reactor system in which the satellite is a core-shell nanoparticle. Hot-carrier generation rates are obtained from an atomistic quantum-mechanical modelling technique which combines a solution of Maxwell's equation with a tight-binding description of the nanoparticle electronic structure. We find that antenna-reactor systems exhibit significantly higher hot-carrier generation rates in the catalytic material than the core-shell system as a result of strong electric field enhancements associated with the gap between the antenna and the satellite. For these systems, we also study the dependence of hot-carrier generation rate on the size of the gap, the radius of the antenna nanoparticle and the direction of light polarization. Our insights pave the way towards a mechanistic understanding of hot-carrier generation in heterogeneous nanostructures for photocatalysis and other energy conversion applications.
\end{abstract}

\maketitle

\section{Introduction}
There is currently significant interest in harnessing energetic or "hot" electrons and holes generated in metallic nanoparticles for applications in photocatalysis~\cite{Robatjazi2017,Mou2012,Liu2019,Yuan2022,Elias}, photovoltaics~\cite{Clavero2014,Enrichi,Salvador2012} and sensing~\cite{goykhman2011locally,li2017harvesting,Chalabi2014,Tang2020,Sun2019,Zhai2019,Zhu2021,faradaytheory}. Metallic nanoparticles feature localized surface plasmons (LSPs) which give rise to large light absorption cross sections~\cite{SMaier}. The LSP has a short lifetime (typically on the order of tens of femtoseconds). Among the various decay mechanisms, the Landau damping decay plays a prominent role because it results in the generation of hot carriers~\cite{Khurgin,Link1999}. 

However, photocatalytic hot-carrier devices often have relatively low efficiencies~\cite{Emi-review}. A possible explanation for this is that standard plasmonic materials, such as Au and Ag, are generally not good catalysts\cite{Hammer1995}. Therefore, attempts have been made to combine plasmonic materials with catalytic materials, such as Pt, Pd or Rh, into functional nanoarchitectures. Examples of such heterostructures include Janus nanoparticles~\cite{Chauhan2018}, core-shell systems~\cite{Rao,Chavez} or nanoparticle dimers and trimers~\cite{Zohar}. Recently, Herran and coworkers studied different nanoarchitectures of Pd and Au, including core-shell nanoparticles and antenna-reactor systems in which a large Au nanoparticle is "decorated" with small Pd nanoparticles (or satellites), for the production of H$_2$ from formic acid~\cite{Herran2022}. These authors found significant enhancements in H$_2$ production upon illumination of the plasmo-catalyst with the largest increase in chemical activity for antenna-reactor systems. Despite these advances, however, there is still no detailed mechanistic understanding of the catalytic activity in bimetallic nanoarchitectures\cite{Zhan2021}.

Insights into microscopic hot-carrier processes, including their generation, thermalization and extraction, can be gained from theoretical modelling. Atomistic first-principles techniques, such as ab initio time-dependent density-functional theory, can be used to investigate hot-carrier processes in very small nanostructures~\cite{Rossi2020}, but are challenging to apply to experimentally relevant system sizes. On the other hand, non-atomistic approaches, such as jellium or spherical well models, can be applied to large systems, but do not capture important aspects, including d-band derived nanoparticle states or facet-specific surface properties~\cite{Manjavacas2014, govorovpapers,lischnergrouppaper1,lischnergrouppaper2,lischnergrouppaper3,lischnergrouppaper4,lischnergrouppaper5}. To address this challenge, Jin and coworkers recently developed a new approach that combines a solution of Maxwell's equation with large-scale atomistic tight-binding models which enables the modelling of hot-carrier processes in nanoparticles containing millions of atoms~\cite{Jin2022}. So far, however, this approach has only been applied to spherical nanoparticles.

In this paper, we use the method of Jin and coworkers~\cite{Jin2022} to study the enhancement hot-carrier generation in a catalytic metal (Pd) induced by the presence of a plasmonic metal (Au) in different bimetallic plasmo-catalytic nanoarchitectures. In particular, we study Au@Pd core-shell nanoparticles and antenna-reactor systems consisting of a large Au nanoparticle which acts as antenna and a small satellite nanoparticle. The satellite is either a spherical Pd nanoparticle or a Au@Pd core-shell nanoparticle. We compare our results to hot-carrier generation rates in spherical Pd nanoparticles. We find that the largest hot-carrier generation rates in the catalytic metal are found in antenna-reactor systems, in particular in those where the satellite nanoparticle is a core-shell system. This can be explained by the large enhancement of the electric field arising from the strongly confined plasmon mode associated with the gap between the antenna nanoparticle and the satellite nanoparticle. We also explore the dependence of hot-carrier generation rates on the light polarization, the size of the antenna nanoparticle and the gap size between the antenna nanoparticle and the satellite. The resulting insights pave the way towards a mechanistic design of heterogeneous nanoarchitectures for energy conversion devices. The approach can readily be applied to other materials.

\section{Methods}
\subsection{Hot-carrier generation rates}
We use the approach developed by Jin and coworkers~\cite{Jin2022} to calculate hot-carrier generation rates in Pd-Au nanoarchitectures. In this method, the rate of hot electrons $N_e(E,\omega)$ of energy $E$ excited by light of frequency $\omega$ is obtained by evaluating Fermi's golden rule according to~\cite{Manjavacas2014}
\begin{equation}\label{eq:Ne}
	N_e\left( E,\omega \right) =\frac{2}{V}\sum_{if}^{} \Gamma_{ if}\left( \omega \right) \delta\left( E-E_f;\sigma \right),
\end{equation} 
where $i$ and $f$ label the initial and final state with energy $E_i$ and $E_f$, respectively. Also, $V$ is the volume of the nanoparticle and we define $\delta(x;\sigma)=\frac{1}{\sqrt{2\pi\sigma^2}}\exp(-\frac{x^2}{2\sigma^2})$ with $\sigma=0.05$ eV being standard deviation of the Gaussian. In the above, $\Gamma_{if}$ is given by 
\begin{equation}
	\Gamma_{if}\left( \omega \right) =\frac{2 \pi}{\hbar}\left|\bra{f} \hat{\Phi}_{tot} \left( \omega \right) \ket{i} \right|^2 \delta(E_f-E_i-\hbar\omega;\gamma)   f\left( E_{ i}  \right)\left( 1-f\left( E_f \right)  \right),
\end{equation}
where $f(E)$ is the Fermi-Dirac distribution with temperature $T=298$~K, $\gamma=0.06$ eV is a broadening parameter and $\hat{\Phi}_{tot}(\omega)$ denotes the total potential inside the nanoparticle. This potential is calculated using the quasistatic approximation \cite{Schnitzer2019,Matias2019,Schnitzer2015,Schnitzer2016,Schnitzer2022,Bohern}.  In particular, we use the finite element method as implemented in COMSOL \textsuperscript{\textregistered}~\cite{comsol} to solve the Laplace equation
\begin{equation}\label{eq:Laplace}
    \nabla\cdot (\epsilon(\mathbf{r},\omega)\nabla \Phi_{tot}(\mathbf{r},\omega))=0, 
\end{equation}
where $\epsilon(\mathbf{r},\omega)$ is dielectric function of the material at position $\mathbf{r}$. We use experimental dielectric functions for Au~\cite{haynes2015crc} and Pd~\cite{haynes2015crc}. In the calculations, we first specify the geometry of the nanoarchitecture and the external potential $\Phi_{ext}(\mathbf{r},\omega)=-\mathbf{E}_0 \cdot \mathbf{r}$ with $\mathbf{E}_0$ denoting the corresponding electric field and then solve Laplace's equation subject to the far-field condition that 
\begin{equation}
    \lim_{|\mathbf{r}|\to \infty}\Phi_{tot}(\mathbf{r},\omega)=\Phi_{ext}(\mathbf{r},\omega).
\end{equation}

Once the total potential of the full Au-Pd nanoarchitecture is determined, we evaluate the hot-carrier generation rate in the Pd subsystem (as only the hot carriers in the Pd are catalytically active). Note that our approach does not capture charge transfer processes between the Au and the Pd which can play an important role in core-shell nanoparticles~\cite{Engelbrekt,Aslam2017}. The electronic states of the Pd subsystem are obtained using the tight-binding method. For this, we first determine the atomic positions by carving the desired Pd shape (either a spherical nanoparticle or a spherical nanoshell) from the bulk material. The eigenstates of the Hamiltonian are expressed in terms of linear combinations of atomic orbitals according to $\ket{i}=\sum_{J,\alpha}C_{J,\alpha}\ket{J,\alpha}$, with $J$ labelling atoms and $\alpha$ labelling orbitals. For each Pd atom, the basis consists of five 4d orbitals, one 5s orbital and three 5p orbitals. The hopping and onsite energies of the Pd tight-binding model are taken from the "Handbook of the Band Structures of Elemental Solids"~\cite{Papaconstantopoulos2015}.

The matrix element in Eq.~\ref{eq:Ne} is evaluated using~\cite{Pedersen2001}
\begin{equation}
    \bra{I,\alpha} \hat{\Phi}_{tot}(\omega) \ket{J,\beta}=\Phi_{tot}(\mathbf{r}_J,\omega)\delta_{I,J}\delta_{\alpha,\beta}, 
\end{equation}
where $\mathbf{r}_J$ denotes the position of atom $J$ and we have ignored the transition dipole contribution to the matrix elements (which we have found to be small in our previous calculations~\cite{Jin2022}).

The total generation rate of hot carriers (both electrons and holes) per unit volume is given by 
\begin{equation}
    N_{tot}(\omega)=2\int_{E_F}^{\infty} N_e(E,\omega)\mathrm{d}E
\end{equation} 
with $E_F$ denoting the Fermi energy. The hot hole generation rate $N_h(E,\omega)$ is obtained by replacing $E_f$ by $E_i$ in Eq.~\ref{eq:Ne}.

\section{Results}
\begin{figure*}
    \centering
    \begin{tikzpicture}
    \node at (0,0) {\includegraphics{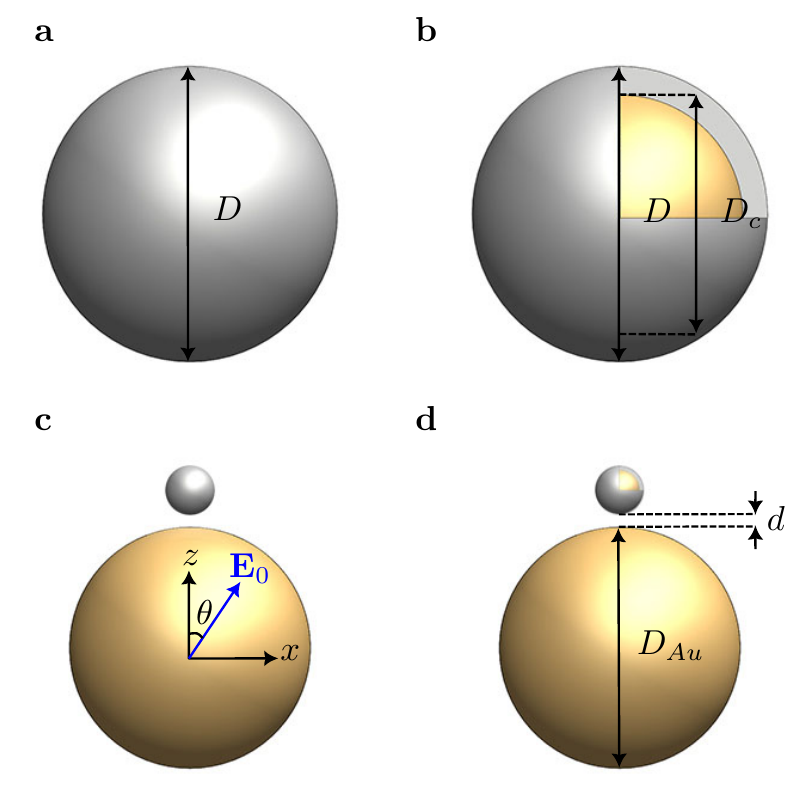}};
    \node at (3*2.54+0.5,-0.5){\includegraphics{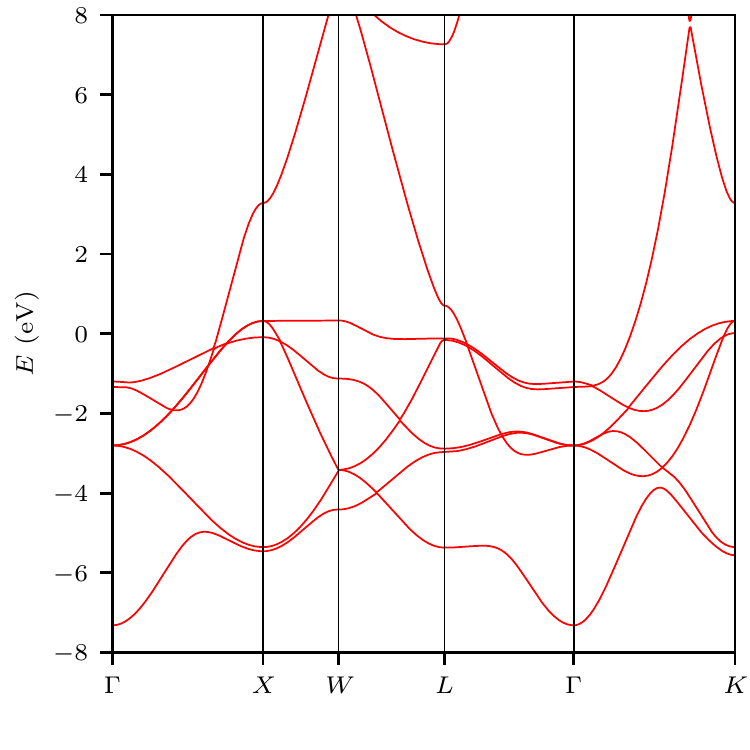}};
    \node at(-3.5+3*2.54+0.5,3.6) {\textbf{e}};
    \end{tikzpicture}
    \caption{Schematic illustration of (a) a spherical Pd nanoparticle, (b) a Au@Pd core-shell nanoparticle, (c) a Au-Pd antenna-reactor system and (d) a Au-Au@Pd antenna-reactor system, (e) band structure of bulk fcc Pd, the zero of energy is set to the Fermi level.}
    \label{fig:systems}
\end{figure*}

We have calculated hot-carrier generation rates in four systems, see Fig.~\ref{fig:systems}: (a) spherical Pd nanoparticles consisting of 3589 atoms (corresponding to a diameter $D=4.42$~nm), (b) spherical Au@Pd core-shell nanoparticles with a Au core (of diameter $D_c=4.90$~nm) and a Pd shell (of thickness 0.40~nm containing 3740 Pd atoms), (c) a Au-Pd antenna-reactor architecture consisting of a small spherical Pd nanoparticle of the same size as in (a) which is separated by a small gap from a larger Au nanoparticle and (d) a Au-Au@Pd antenna-reactor architecture in which the small nanoparticle has a core-shell structure as in (b). The sizes of the Pd nanoparticle and of the Au@Pd core-shell nanoparticle are similar to the satellite nanoparticles used in the experiment of Herran and coworkers~\cite{Herran2022}.

\begin{figure*}
    \centering
    \begin{tikzpicture}
    \node at (0,0) {\includegraphics{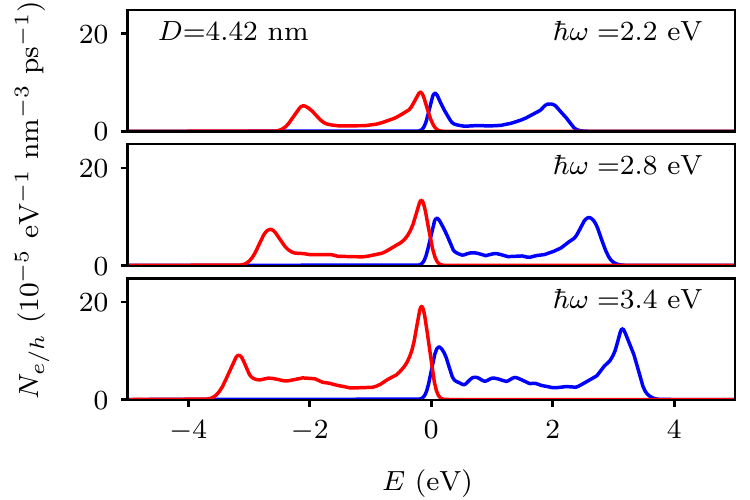}};
    \node at (3*2.54,0){\includegraphics{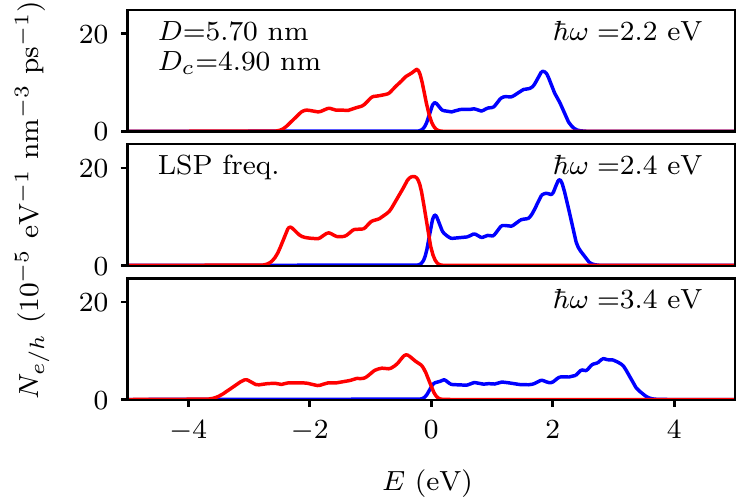}};
    \node at(-3.5,2.7) {\textbf{a}};
    \node at(-3.5+3*2.54,2.7) {\textbf{b}};
    \end{tikzpicture}
    \caption{Dependence of hot-carrier generation rate on photon energy for (a) a spherial Pd nanoparticle with a diameter of $D=4.42$~nm and (b) a Au@Pd core-shell nanoparticle with a core diameter of $D_c=4.90$~nm and a shell thickness of 0.40 nm. Hot hole (electron) generation rates are red (blue). The zero of energy is set to the Fermi level.}
    \label{fig:Palladium_and_coreshell}
\end{figure*}

Figure~\ref{fig:Palladium_and_coreshell}(a) shows the evolution of the hot carrier generation rate for the spherical Pd nanoparticle as function of photon energy. Both the hot hole and the hot electron rates exhibit two peaks. The hot hole rate has a peak near the Fermi level with the hot electron rate having a corresponding peak at $\sim \hbar \omega$ above the Fermi level. The other peak of the hot hole rate is at $\sim -\hbar \omega$ and the hot electron rate has a corresponding peak just above the Fermi level. As the photon energy is varied, the peaks near the Fermi level are pinned at their positions, but the other peaks move to higher (in case of the hot electron rate) and lower (in case of the hot hole rate) energies. This finding can be understood from an analysis of the electronic structure of Pd. In particular, the band structure of Pd, see Fig.~\ref{fig:systems}(e), exhibits flat d-bands just above and below the Fermi energy. These flat bands give rise to a large density of states which translates into a high hot carrier generation rate near the Fermi level. The positions of the other peaks are then determined by energy conservation, i.e. for each electron of energy $E$ a corresponding hole of energy $E-\hbar \omega$ must be created. Finally, we do not observe a dramatic enhancement of the hot-carrier rates at a particular photon energy which reflects the absence of a strong LSP resonance in spherical Pd nanoparticles. 

Figure~\ref{fig:Palladium_and_coreshell}(b) shows the dependence of the hot carrier generation rate of a Au@Pd core-shell nanoparticle on the photon energy. Again, the rates exhibit peaks near the Fermi level. Some differences in the shapes of the hot-carrier generation rates can be observed compared to the spherical Pd nanoparticle: these are caused by the increase of surface area of the thin Pd shell which enhances the generation of hot carriers from intraband transitions~\cite{Jin2022}. In the core-shell system, an enhancement of the hot-carrier generation rate at a photon frequency of 2.4 eV can be observed. This energy is close to the LSP energy of a spherical Au nanoparticle. In other words, at this photon energy the field enhancement of the Au core increases the hot-carrier generation in the Pd shell~\cite{Aslam2017}.

\begin{figure*}
    \centering
    \begin{tikzpicture}
    \node at (0,0) {\includegraphics{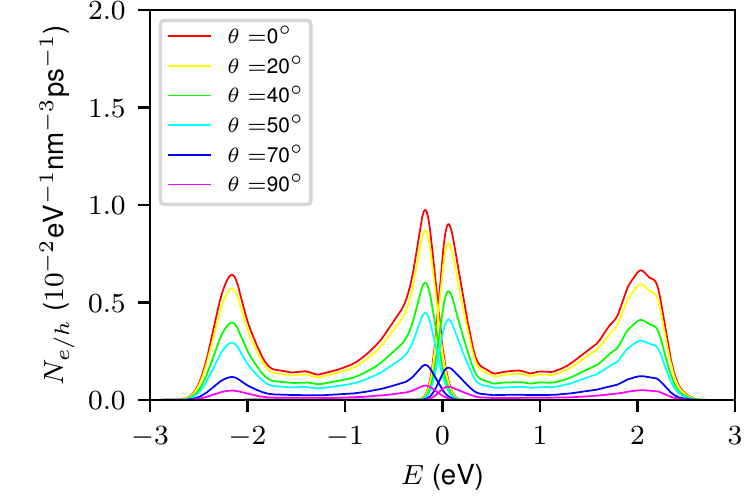}};
    \node at (3*2.54,0){\includegraphics{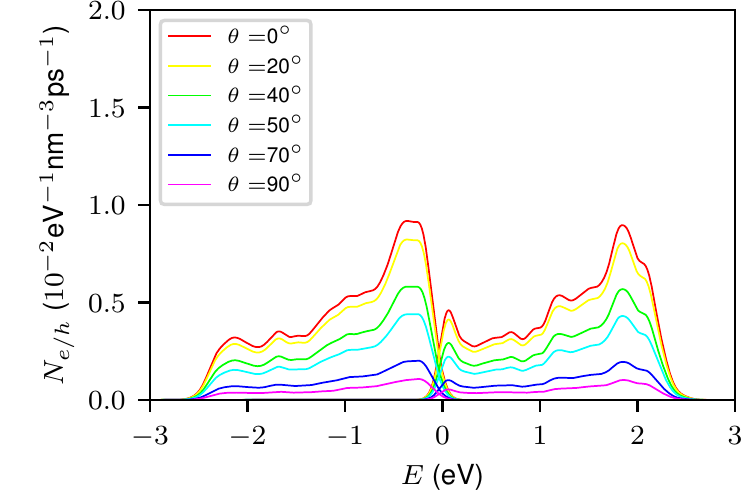}};
    \node at(-3.5,2.7) {\textbf{a}};
    \node at(-3.5+3*2.54,2.7) {\textbf{b}};
    \end{tikzpicture}
    \caption{Dependence of hot-carrier generation rate on light polarization for (a) a Au-Pd antenna-reactor system with a diameter of 4.42 nm, (b) Au-Au@Pd reactor system, with a core diameter of 4.9 nm and a total diameter of 5.7 nm. The diameter of the Au nanoparticle is 49 nm. The size of the gap between the Pd and Au nanoparticles is 0.40~nm and the photon energy is is 2.28~eV for Au-Pd system and 2.20 for Au-Au@Pd system.}
    \label{fig:angle_dependence}
\end{figure*}

Next, we investigate hot-carrier generation in the Au-Pd antenna-reactor system. Adding the Au nanoparticle to the spherical Pd nanoparticle lifts the rotational symmetry of the Pd system. As a consequence, the hot-carrier generation rate now depends on the polarization vector of the electric field. We assume that both the centre of the Au and the centre of the Pd nanoparticle lie on the z-axis, see Fig.~\ref{fig:systems}, and describe the electric field through its polar angle $\theta$. Fig.~\ref{fig:angle_dependence}(a) shows the dependence of the hot carrier rate on $\theta$. The hot-carrier generation rate is largest when $\theta=0^\circ$ and then decreases as $\theta$ is increased. In particular, we find that the hot-carrier generation rate is proportional to $\cos\theta$, i.e. $N_e(\omega,E,\theta)=N_e(\omega,E,0)\cos\theta$. When $\theta=0$, the charge carriers in the Au nanoparticle are pushed towards the Pd nanoparticle creating a strongly enhanced electric field. The field is further enhanced by the confinement effect of the small gap between between the Au and the Pd nanoparticles giving rise to a so-called gap plasmon~\cite{gapplasmon}. We note that a large number of satellites is attached to the Au nanoparticles in the experiments of Herran and coworkers~\cite{Herran2022}. Also, the light used in experiments is not polarized. Therefore, the experimentally measured hot-carrier generation rate is an average over $\theta$.

Figure~\ref{fig:angle_dependence} shows the same results for a Au-Au@Pd antenna-reactor system. Again, the hot-carrier generation rate of the core-shell system has a different shape than the pure Pd system, but it exhibits a similar dependence on the light polarization.

\begin{figure*}
    \centering
    \begin{tikzpicture}
    \node at (0,0) {\includegraphics{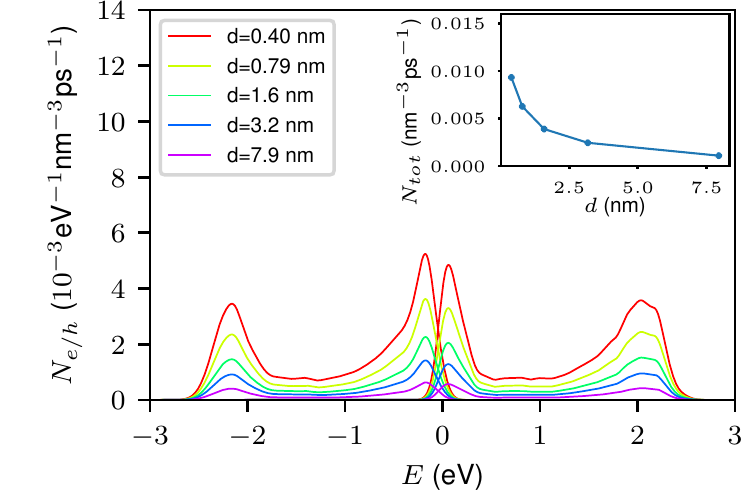}};
    \node at (3*2.54,0){\includegraphics{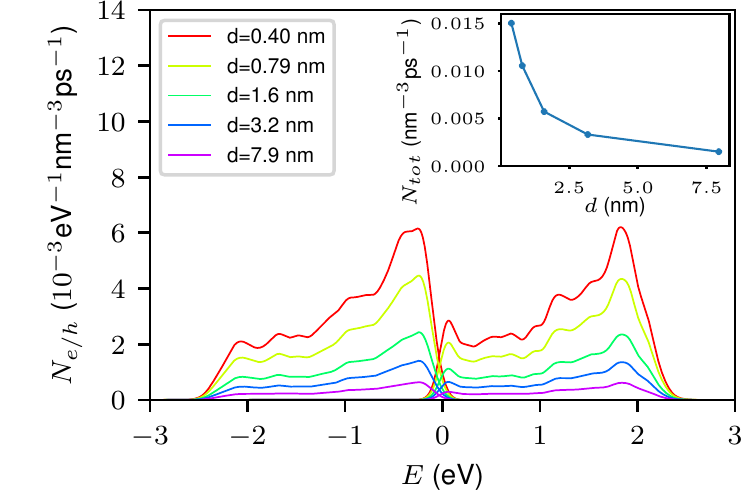}};
    \node at(-3.5,2.7) {\textbf{a}};
    \node at(-3.5+3*2.54,2.7) {\textbf{b}};
    \end{tikzpicture}
    \caption{Dependence of the hot-carrier generation of Au-Pd antenna-reactor systems on the size of the gap $d$ between the Au nanoparticle and the satellite. (a) Results for a Pd satellite. (b) Results for a Au@Pd core-shell satellite. The diameter of the Au antenna nanoparticle 49~nm, the photon energy is 2.28~eV for Au-Pd system and 2.20 for Au-Au@Pd system and the calculation is averaged over all polarization angles. The insets show the total hot-carrier generation rate}
    \label{fig:distance_dependence}
\end{figure*}

Next, we study the dependence of the hot-carrier generation rate of the two antenna-reactor systems on the size of the gap $d$ between the Au nanoparticle and the satellite, see Fig.~\ref{fig:distance_dependence}. In practice, this parameter is determined by the size of the ligands on the surface of the nanoparticles and is difficult to control experimentally (in the experiment of Herran et al., the gap is estimated to be approximately 1 nm~\cite{Herran2022}). We find that the largest hot-carrier rates are obtained for the systems with the smallest gaps. The insets show that the total number of hot carriers decreases quickly as the gap size is increased. For example, increasing the gap from $0.49$~nm to $1.6$~nm reduces the total number of hot carriers by a factor of 0.38 (Au-Au@Pd) and 0.42 (Au-Pd) and further increase to $d=3.2$~nm gives rise to an additional reduction by a factor of 0.58 (Au-Au@Pd) and 0.63 (Au-Pd) in the total hot carrier generation rate. This result demonstrates the importance of the confinement effect associated with the gap between the Au and the satellite particles on the strength of the electric field.

\begin{figure*}
    \centering
    \begin{tikzpicture}
    \node at (0,0) {\includegraphics{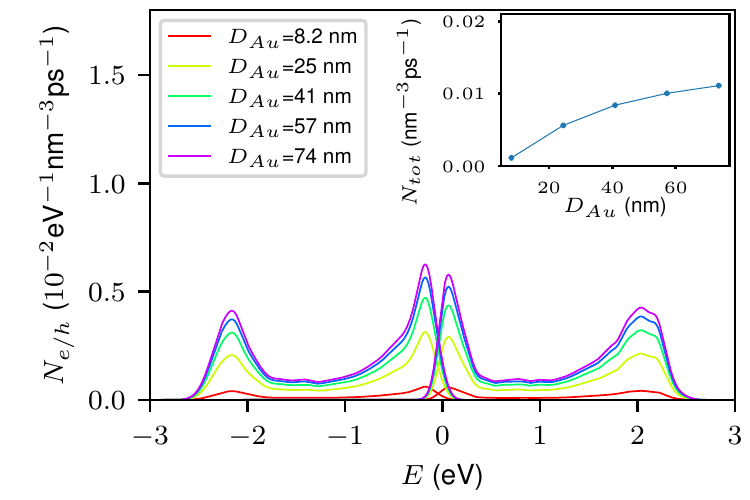}};
    \node at (3*2.54,0){\includegraphics{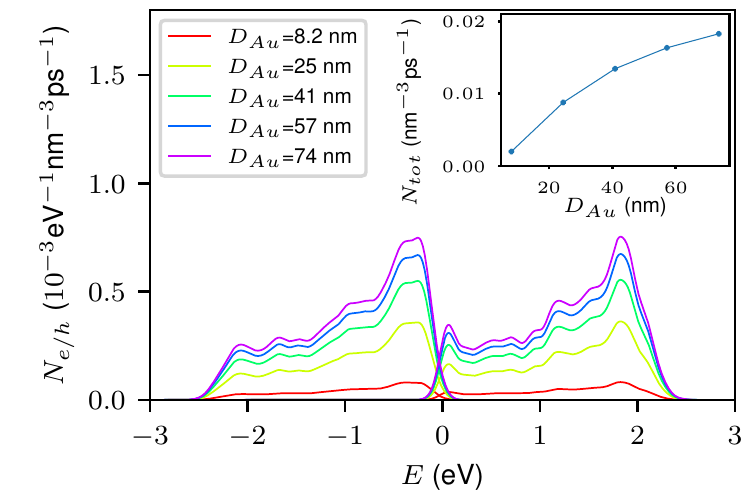}};
    \node at(-3.5,2.7) {\textbf{a}};
    \node at(-3.5+3*2.54,2.7) {\textbf{b}};
    \end{tikzpicture}
    \caption{Dependence on the hot-carrier generation of Au-Pd antenna-reactor systems on the size of the Au nanoparticle. (a) Results for a Pd satellite. (b) Results for a Au@Pd core-shell satellite. The size of the gap is 0.40~nm, the photon energy is 2.28~eV for Au-Pd system and 2.20~eV for Au-Au@Pd system and the calculation was averaged over all polarization angles. The insets show the total hot-carrier generation rates.}
    \label{fig:size_dependence}
\end{figure*}

Next, we study the dependence of the Pd hot-carrier generation rate of the reactor systems on the size of the Au nanoparticle, see Fig.~\ref{fig:size_dependence}.
We find that the hot-carrier generation rate increases as the size of the Au nanoparticle increases. Increasing the Au nanoparticle size while keeping the distance between the Au nanoparticle and the satellite fixed reduces the volume available in the gap between the two nanoparticles and thus enhances confinement effects. This in turn leads to electric field enhancement. As the Au nanoparticle size increases, the additional reduction of the gap volume becomes less and less and only results in a small increase in hot carrier production.

\begin{figure*}
    \centering
    \begin{tikzpicture}
    \node at (0,0) {\includegraphics{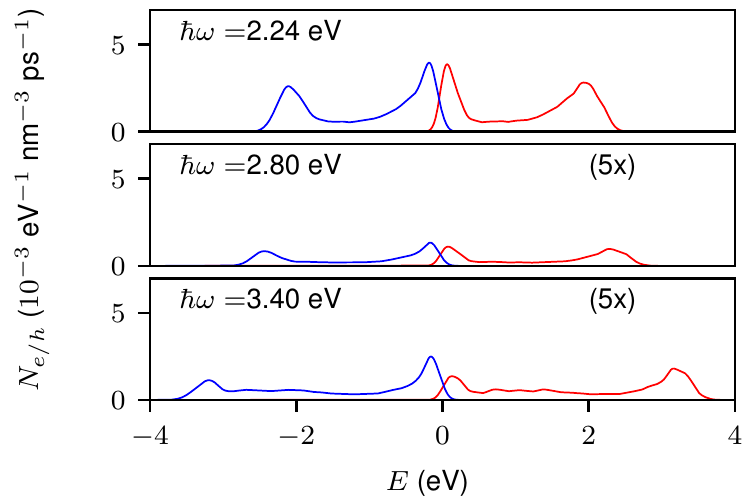}};
    \node at (3*2.54,0){\includegraphics{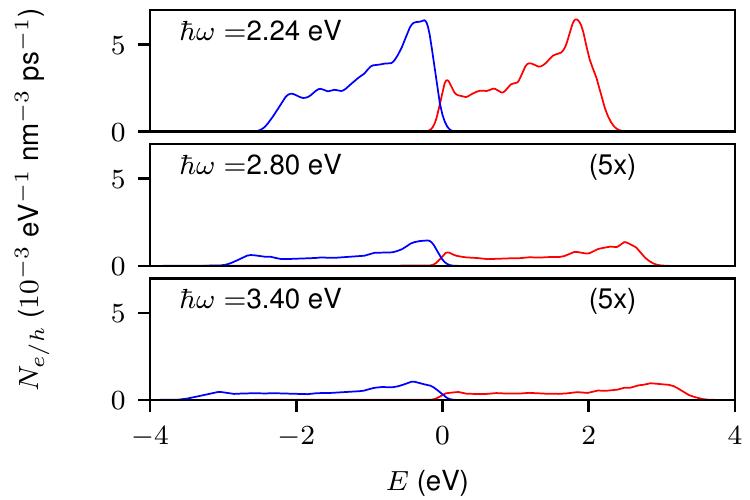}};
    \node at(-3.5,2.7) {\textbf{a}};
    \node at(-3.5+3*2.54,2.7) {\textbf{b}};
    \end{tikzpicture}
    \caption{Dependence of hot-carrier generation rate on the photon energy for (a) a Au-Pd antenna-reactor system with a diameter of 4.42 nm, (b) Au-Au@Pd antenna-reactor system, with a core diameter of 4.9 nm and a total diameter of 5.7 nm. The diameter of the Au nanoparticle is 49 nm. The size of the gap between the Pd and Au nanoparticles is 0.40~nm, and the calculation is averaged over all polarisation angles. For visual aid, the hot-carrier generation rates at $\hbar\omega=2.80$~eV and 3.40 eV were multiplied by a factor of 5.}
    \label{fig:frequency_dependence}
\end{figure*}

Finally, we study the dependence of the hot-carrier generation rates of the reactor systems on the photon energy, see Fig.~\ref{fig:frequency_dependence}. For both systems, a dramatic enhancements of the rates is observed at a photon energy of 2.24 eV compared to the other photon energies. This photon energy is close to the LSP energy of the Au nanoparticle (note that the presence of the satellite modifies the LSP energy of the Au nanoparticle) so the increase of the hot-carrier generation rates reflects the electric field enhancement caused by the plasmon mode. The LSP acts as an optical lens which efficiently funnels energy towards the catalytic material.

\begin{figure*}
    \centering
    \begin{tikzpicture}
    \node at (0,0) {\includegraphics{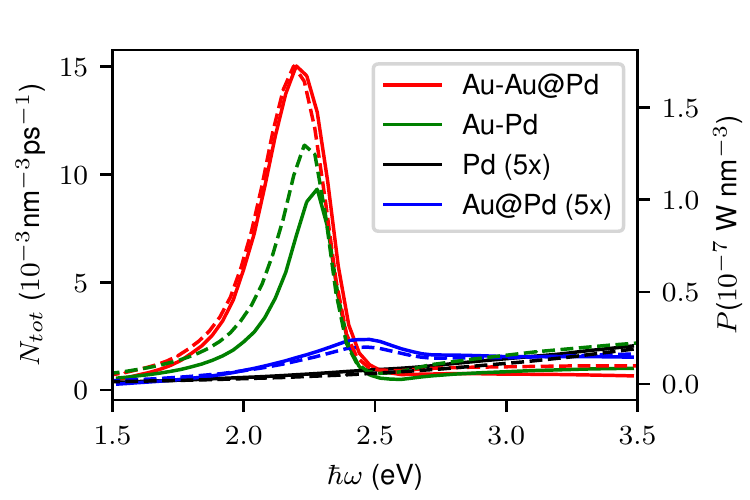}};
    \node at (3*2.54+0.5,-0.4){\includegraphics{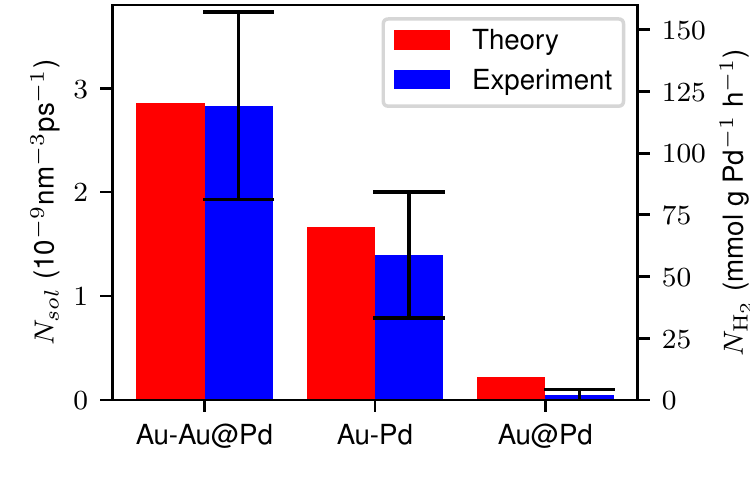}};
    \node at(-3.5,2.7) {\textbf{a}};
    \node at(-3.5+3*2.54,2.7) {\textbf{b}};

    \end{tikzpicture}
    \caption{(a) Total hot carrier generation rate as a function of photon energy (averaged over all polarisation vectors) for a spherical Pd nanoparticle, a Au@Pd core-shell nanoparticle, a Au-Pd reactor and a Au-Au@Pd reactor. The power (per unit volume) absorbed by the Pd subsystem is shown in dashed lines. Note that the curves for the Pd nanoparticles and the Au@Pd have been multiplied by a factor of 5 to improve visibility. (b) Comparison of the measured difference in H$_2$ production upon illumination by a solar simulator and in the dark to the calculated number of hot holes excited by solar illumination.}
    \label{fig:four_structures}
\end{figure*}

We compare the total Pd hot-carrier generation rates of the 4 different systems (spherical Pd nanoparticle, Au@Pd core-shell nanoparticle, Au-Pd antenna-reactor system and Au-Au@Pd antenna-reactor system) in Fig.~\ref{fig:four_structures}(a). It can be seen that the two antenna-reactor systems produce significantly more hot carriers in the Pd than the other two systems. In particular, a dramatic increase in the generation rate is observed near the plasmon frequency of $\sim 2.2$~eV. As discussed above, the large generation rates are a consequence of the gap plasmon mode, which gives rise to large electric field enhancements. In the antenna-reactor system with a core-shell satellite, the electric field is more strongly confined because of the presence of the Au inside the core-shell satellite and therefore this system gives rise to the largest hot-carrier generation rate overall. 

In their experiments, Herran and coworkers measure the increase in H$_2$ production from formic acids by bimetallic plasmonic catalysts upon illumination with a solar simulator (see Figure 2(a) of Ref.~\cite{Herran2022}). In Fig.~\ref{fig:four_structures}(b) we show the measured difference in H$_2$ production in the dark and upon illumination for a Au@Pd core-shell nanoparticle, a Au-Pd antenna-satellite system and a Au-Au@Pd system. We compare this difference to the total number of hot holes $N_{sol}$ generated by the solar spectrum $S(\omega)$ (using the Air Mass 1.5 spectrum) obtained via
\begin{equation}
    N_{sol} = \int d\omega S(\omega) N_{tot}(\omega).
\end{equation}
In excellent agreement with experiment, we find that the Au-Au@Pd antenna-satellite system is the best bimetallic plasmonic photocatalyst while the Au@Pd system performs much worse.

\section{Conclusion}
We have studied hot-carrier generation in Au-Pd nanoarchitectures using an atomistic quantum-mechanical modelling approach. We have found that Au-Pd antenna-reactor systems exhibit significantly higher hot-carrier generation rates than core-shell nanoparticles. This is caused by the large electric field enhancements due to the localized plasmon mode associated with the gap between the antenna and the satellite nanoparticles. In particular, the largest overall hot-carrier generation rate is found for an antenna-reactor system in which the satellite is a core-shell nanoparticle. For the antenna-reactor systems, we also studied the dependence of the hot-carrier generation rates on the size of the gap, the radius of the antenna nanoparticle and the light polarization direction. We find that the largest rates are found when the electric field is parallel to the axis connecting the centres of the antenna and satellite nanoparticles. Also, small gaps and large antenna sizes favor hot-carrier generation. The insights from our work can guide the development of highly efficient heterogeneous hot-carrier nanodevices for energy conversion applications.

\section{Acknowledgments} 
HJ acknowledges financial support from his parents. M.H. and E.C. acknowledge the Deutsche Forschungsgemeinschaft (DFG, German Research Foundation) under e-conversion Germany´s Excellence Strategy – EXC 2089/1 – 390776260, the Bavarian program Solar Energies Go Hybrid (SolTech), the Center for NanoScience (CeNS) and the European Commission for the ERC-STG Catalight (802989). JL acknowledges funding from the EPSRC programme grant EP/W017075/1.

\end{document}